\documentclass[letterpaper, 10 pt, conference]{ieeeconf}  

\IEEEoverridecommandlockouts                              
 \usepackage{amssymb}
\usepackage{graphicx}
\usepackage{array}
\usepackage{mathrsfs}
\usepackage{amsfonts}
\usepackage{setspace}
\usepackage{amsmath}
\usepackage{amssymb}
\usepackage{mathrsfs}
\usepackage{subfigure}

\input xy
\xyoption{all}

\newtheorem{theorem}{Theorem}
\newtheorem{lemma}{Lemma}

\newtheorem{definition}{Definition}

\newtheorem{problem}{Problem}

\title{\LARGE \bf
Decentralized Hybrid Formation Control \\of Unmanned Aerial Vehicles
}

\author{Ali Karimoddini$^{1}$,  Mohammad Karimadini $^{2}$, Hai Lin$^{3}$
\thanks{$^{1}$  A. Karimoddini is with the Department of  Electrical and Computer Engineering, North Carolina Agricultural and Technical State University, Greensboro, NC 27411 USA, {\tt\small akarimod@ncat.edu}.
        }%
\thanks{$^{2}$M. Karimadini is with the Department of Electrical Engineering, Arak University of Technology, Arak, Iran, {\tt\small elekm@nus.edu.sg}.
            }%
\thanks{$^{3}$H. Lin is with the Department of Electrical Engineering,
University of Notre Dame, Notre Dame, USA, {\tt\small hlin1@nd.edu}.
               }%
}

\begin{document}

\maketitle
\thispagestyle{empty}
\pagestyle{empty}

\begin{abstract}
 This paper presents a  decentralized hybrid supervisory control approach for a team of unmanned helicopters that are involved in a leader-follower formation mission. Using a polar partitioning technique,  the motion dynamics of the follower helicopters are abstracted to  finite state machines. Then, a
discrete supervisor is designed in a modular way for different components of the formation mission including reaching the formation, keeping the formation, and collision avoidance. Furthermore, a formal  technique is developed to design the local supervisors decentralizedly, so that the team of helicopters as whole, can cooperatively accomplish a collision-free formation task.
\end{abstract}

\section{INTRODUCTION}
Nowadays,  developing Unmanned Aerial Vehicles (UAVs) in different sizes and shapes for various applications has
emerged as an attractive research area \cite{valavanis2007advances}, \cite{bortoff1999university},
\cite{michelson1998update},    \cite{partovi2012development}. A challenging problem in the aerial robotics area and cooperative control of UAVS is \emph{formation control}, in which it is desired  to instruct a group of agents to jointly move with a relatively fixed distance. This capability improves the performance of UAVs to accomplish different tasks such as search and coverage more efficiently.  In the literature, there are several methods that can partly handle subcomponents of a formation mission.
 For instance, for \emph{reaching the formation}, methods such  as MILP programming, navigation function, and potential field
have been developed \cite{980728, Koditschek:1990:RNF:95086.95100, 1387478, 1023918}. \emph{Keeping the formation} can be seen as a standard
control problem in which the system's actual position has slightly
deviated from the desired position \cite{Stipanovic20041285, 4136907, Giulietti200565}. Finally, in \cite{4655013, 4602109, 4407015, 1428839},
different scenarios for \emph{collision avoidance} have been introduced
using geometry approaches,
 predictive control, probabilistic methods, and  invariant sets.
 Nevertheless, putting all together to address the whole components of the formation mission, requires an  in-depth understanding of the interplay between the components based on which a decision making unit can be embedded in the control structure of  the UAVs. To make this control structure reliable enough, two main problems should be addressed. Firstly, this control structure has a hybrid nature, which includes both the continuous dynamics of the UAVs and  the discrete dynamics of the decision making unit that interactively coexist in the system
\cite{antsaklis1993hybrid}. Although a common practice is to treat the continuous and the discrete structure of the system in a decoupled way, the ignorance of the interactions between the continuous and discrete dynamics of the system degrades the reliability of the overall system. Secondly, to take the advantage of decentralized control schemes, e.g.  distributing  the computation costs among the agents and  increasing the reliability of the system against the possible failures, a decentralized controller is required. To address the first problem, in  \cite{myformationmechatronics}, a hybrid supervisory control framework  was introduced for the formation control of UAVs.

This paper addresses the second problem and presents a decentralized hybrid supervisory control of UAVs that are involved in a leader-follower formation scenario. First, using the abstraction techniques,  a DES model is obtained for the motion dynamics of each agent. Then, the formation task is formulated by logical requirements for which we have modularly designed the discrete supervisors for different components of the formation including reaching the formation, keeping the formation, and collision avoidance.  In the reaching and keeping the formation, the follower UAVs can satisfy the desired performance independently. However, for the collision avoidance, a tight cooperation of the UAVs is required. For this purpose, a collision avoidance supervisor is designed, so that the team of UAVs as whole, can cooperatively satisfy the collision avoidance specification as a global goal. Then, to render the decentralized implementation, the designed global supervisor is decomposed into  local supervisors through the natural projections into local event sets.

The rest of this paper is organized as follows. Section \ref{PROBLEM FORMULATION} describes the problem  formulation.  Section \ref{Discrete model of the UAV motion dynamics over the partitioned space} obtains an abstract model for the motion dynamics of the follower UAVs using the polar partitioning of the motion space. A discrete supervisor is modularly designed in Section \ref{Decentralized Modular supervisor design for the formation control of the UAVs},  and then, it is decomposed  into
 local supervisors.
The paper is concluded in Section \ref{CONCLUSION}.

\section{Problem formulation}\label{PROBLEM FORMULATION}
In \cite{intechbook} and \cite{karimoddini2010multi} it is shown that subject to the proper implementation of the inner-loop for an unmanned helicopter to be fast enough to track the given
references, the outer loop dynamics can be
approximately described as follows:

\begin{equation}\label{dynamics}
    \dot{x}=u, \,\,\,\,\,x\in \mathbb{R}^2,\,\,\,\,\,u\in U\subseteq
    \mathbb{R}^2,
\end{equation}

 where $x$ is the position of the UAV; $u$ is the UAV velocity
 reference generated by the formation algorithm, and $U$ is the convex set of velocity
 constraints.


 Also, assume that the UAVs are flying at the same altitude,
and the velocity of the k'th follower, $UAV_k$, $k=1,2$ is in the following form:

   \begin{equation}\label{rel}
{V_{follower}}_k=V_{leader}+{V_{rel}}_k.
\end{equation}

Now, we can consider
a relatively fixed frame for each follower UAV, in
which each follower moves with the relative velocity $V_{rel}$.

\begin{problem}\label{problem1}
\emph{Given the dynamics of the follower UAVs as (\ref{dynamics}) and their
velocity  in the form of (\ref{rel}), design the formation
controller to generate the relative velocity of the followers,
${V_{rel}}_k$, such that starting from any initial state inside the
control horizon, the follower UAVs eventually reach their desired positions, while
avoiding the collision with other follower UAVs.
Moreover, after reaching the formation, the follower UAVs should
remain at the desired positions.}
\end{problem}


\section{Discrete model of the UAV motion dynamics over the polar partitioned space}\label{Discrete model of the UAV motion dynamics over the partitioned space}
To address this problem, for each UAV consider a circle with the radius of $R_m$  that is centered at its
desired position. With the aid of the   partitioning
curves $\{ r_i = \frac{{R_m }}{{n_r }-1}(i-1),\,\,i = 1,...,n_r \}$
and $\{ \theta _j = \frac{2\pi }{{n_\theta }-1}(j-1),\,\,j =
1,...,n_\theta \}$, this circle can be partitioned into
$(n_r-1)(n_{\theta}-1)$ partitioning elements.

\begin{figure}
\centering 
\subfigure[] 
{
  \includegraphics[width=1.5 in, height=1.2in]{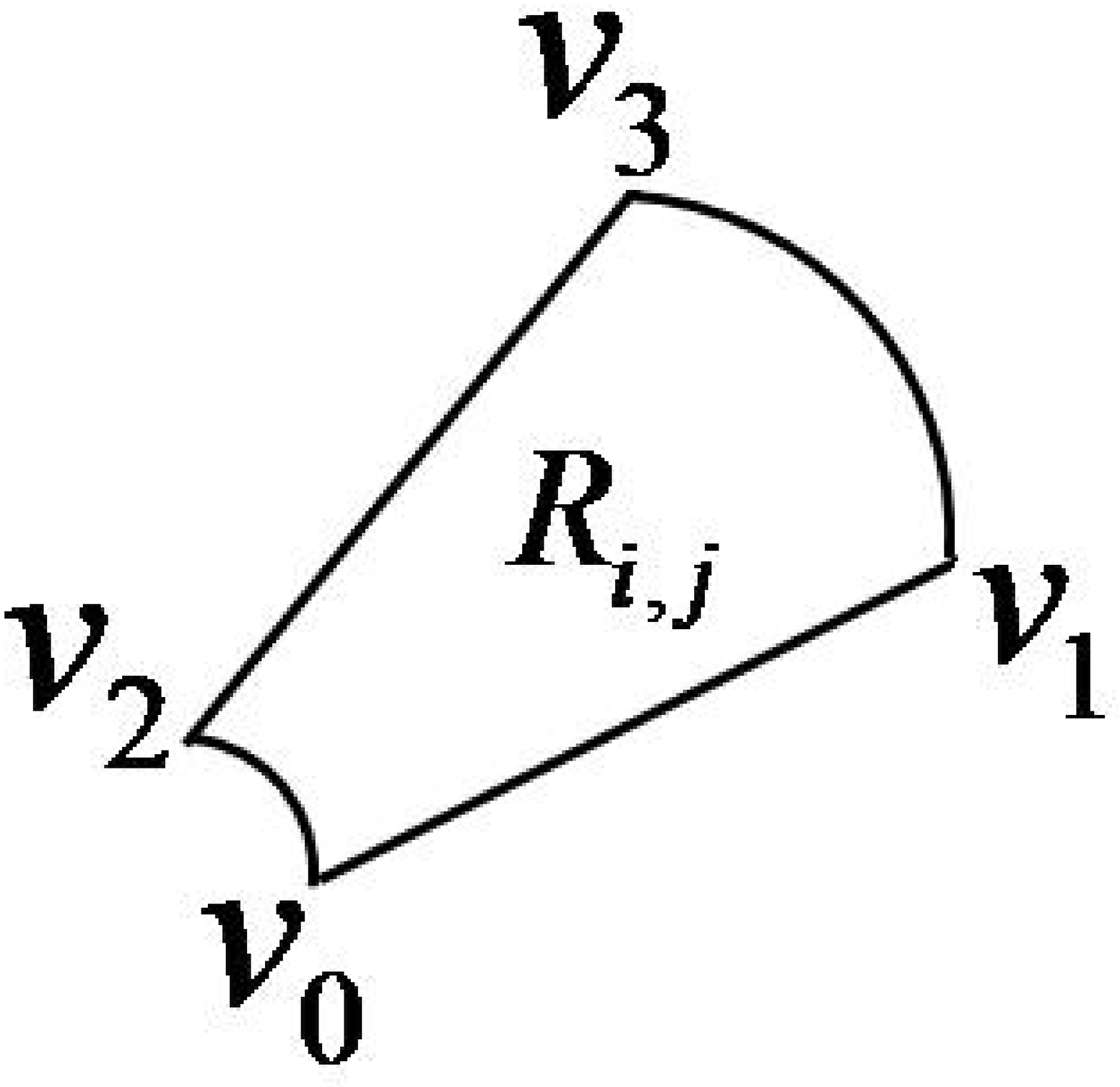}
  \label{fig4epsvertices}
}
\subfigure[] 
{
  \includegraphics[width=1.5in, height=1.2in]{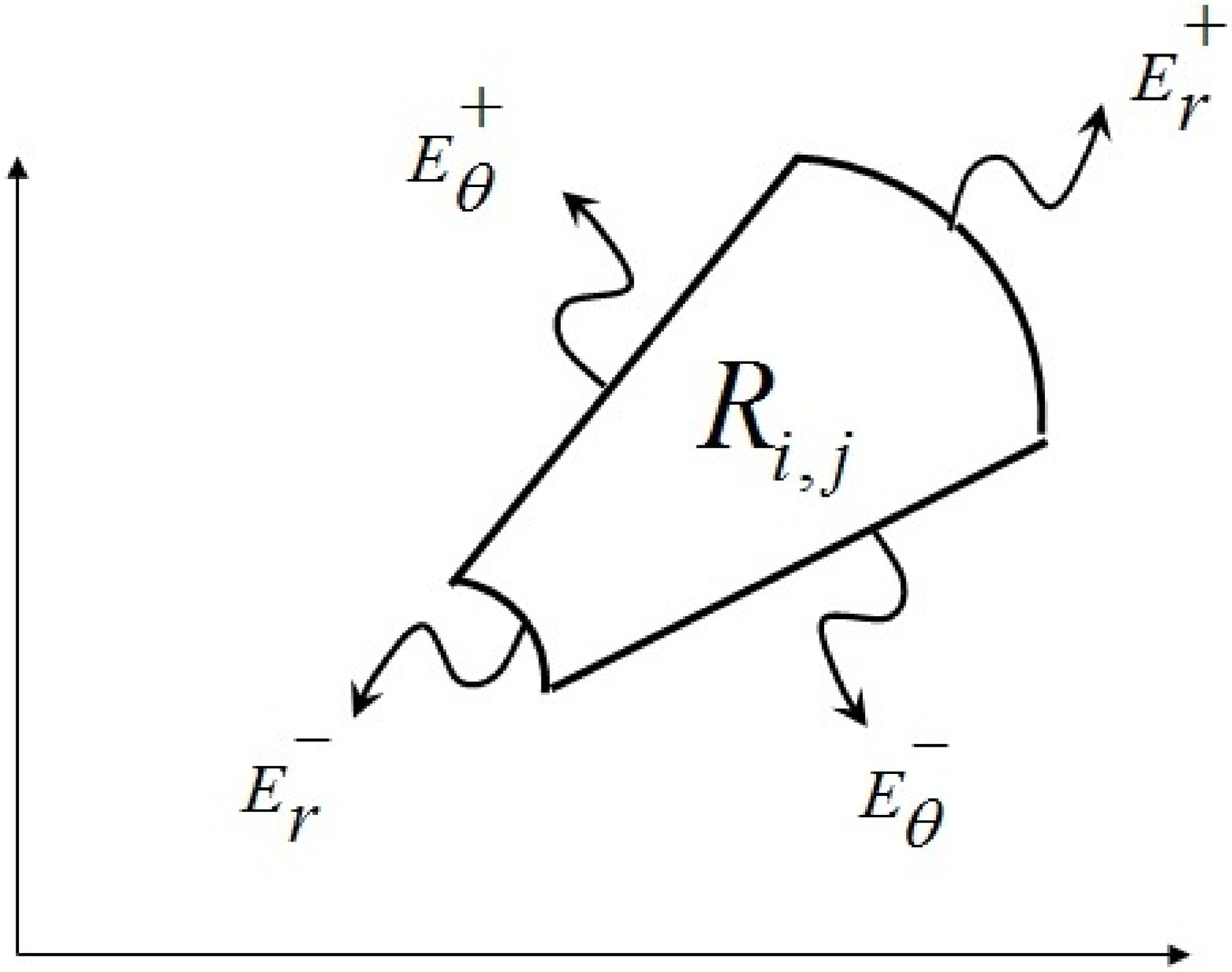}
  \label{edges}
}
\caption{(a) Vertices of the element $R_{i,j}$. (b) Edges of the element $R_{i,j}$. } 
\label{fig5epsouternormals}
\end{figure}

In this partitioned space, an element $R_{i,j}=\{p=(r,\theta)|\, r_{i}\leq
r\leq r_{i+1},\,\theta_{j}\leq\theta\leq\theta_{j+1}\}$, has four
vertices, $v_0,v_1,v_2,v_3$ (Fig. \ref{fig4epsvertices}), four edges, $E_r^+$, $E_r^-$, $E_{\theta}^+$, $E_{\theta}^-$ (Fig. \ref{edges}). The set $V(\ast)$ stands for the vertices that belong to
 $\ast$ ($\ast$ can be an edge, or a region ${R}_{i,j}$).

As shown in \cite{myformationmechatronics},  for  a system with a multi-affine dynamics $\dot x =
h(x,u(x))$ defined over this polar partitioned space, two control features can be designed.  First,  the region $R_{i,j}$ can be  invariant, i.e., the trajectories of  the system remain inside the region forever.
 The other control feature is the exit edge. It is possible to design a controller to drive the system's trajectory to exit from the edge $E_q^s$, $q \in \{ r,\theta \}$ and $s
\in \{ + , - \}$, by choosing the control values $u(v_m)$ at the vertices.
According to the properties of multi-affine systems,  the control value at any point inside the region can be achieved based on the control values at the vertices as $u(x)= \Sigma_{m=0}^3
{\lambda _m(x) u(v_m )}$, which $\lambda _m(x)$ is a coefficient that determines the weight of $u(v_m )$ in the control value $u(x)$. We denote the controller for having a region invariant  by ${C_0}_k$. Also, ${C_r}_k^ +$, ${C_r}_k^ -$, ${C_{\theta }}_k^ +$, and ${C_{\theta }}_k^-$ are respectively the controllers for having the edges
${F_r}^ +$, ${F_r}^ -$, ${F_{\theta }}^ +$, and ${F_{\theta }}^-$, as exit edges. Further details on how to design these controllers are provided in \cite{myformationmechatronics}.

Now, this model of the UAV motion dynamics over the partitioned space can be abstracted to a finite state machine and can be presented by  a discrete automaton. An automaton can be formally defined as follows:
\begin{definition}(Automaton)\label{Automaton}\cite{Cassandras2008}.
A deterministic automaton is a tuple $A := \left(Q, q_0, E, \delta,Q_m
\right)$ consisting of a set of states $Q$; an initial state $q_0\in
Q$; a set of events $E$ that causes transitions between the states,
and a transition relation $\delta \subseteq Q \times E\times Q$
(with a partial map $\delta: Q \times E \to Q$), such that $(q, e,
q^{\prime})\in \delta$ if and only if state $q$ is transited to
state $q^{\prime}$ by event $e$, denoted by
$q\overset{e}{\underset{}\rightarrow } q^{\prime}$ (or $\delta(q, e)
= q^{\prime}$). $Q_m
\subseteq Q$ represents the marked states to assign a
meaning of accomplishment to some states. For supervisor automaton whose
all states are marked, $Q_m$ is omitted from the tuple.
\end{definition}

For this automaton, the sequence of these
events forms a string.
 We use $\varepsilon$ to denote an empty string, and $\Sigma ^{*}$ to denote the set of
all possible strings over the set $\Sigma$ including $\varepsilon$.
The language of the automaton $G$, denoted by $L(G)$, is the set of
all strings that can be generated by $G$, starting from the initial
states. The marked language, $L_m(G)$, is the set of strings that
belong to  $L(G)$ and end with the marked states.

\begin{figure}
\centering 
\subfigure[] 
{
  \includegraphics[width=1 in, height=.75in]{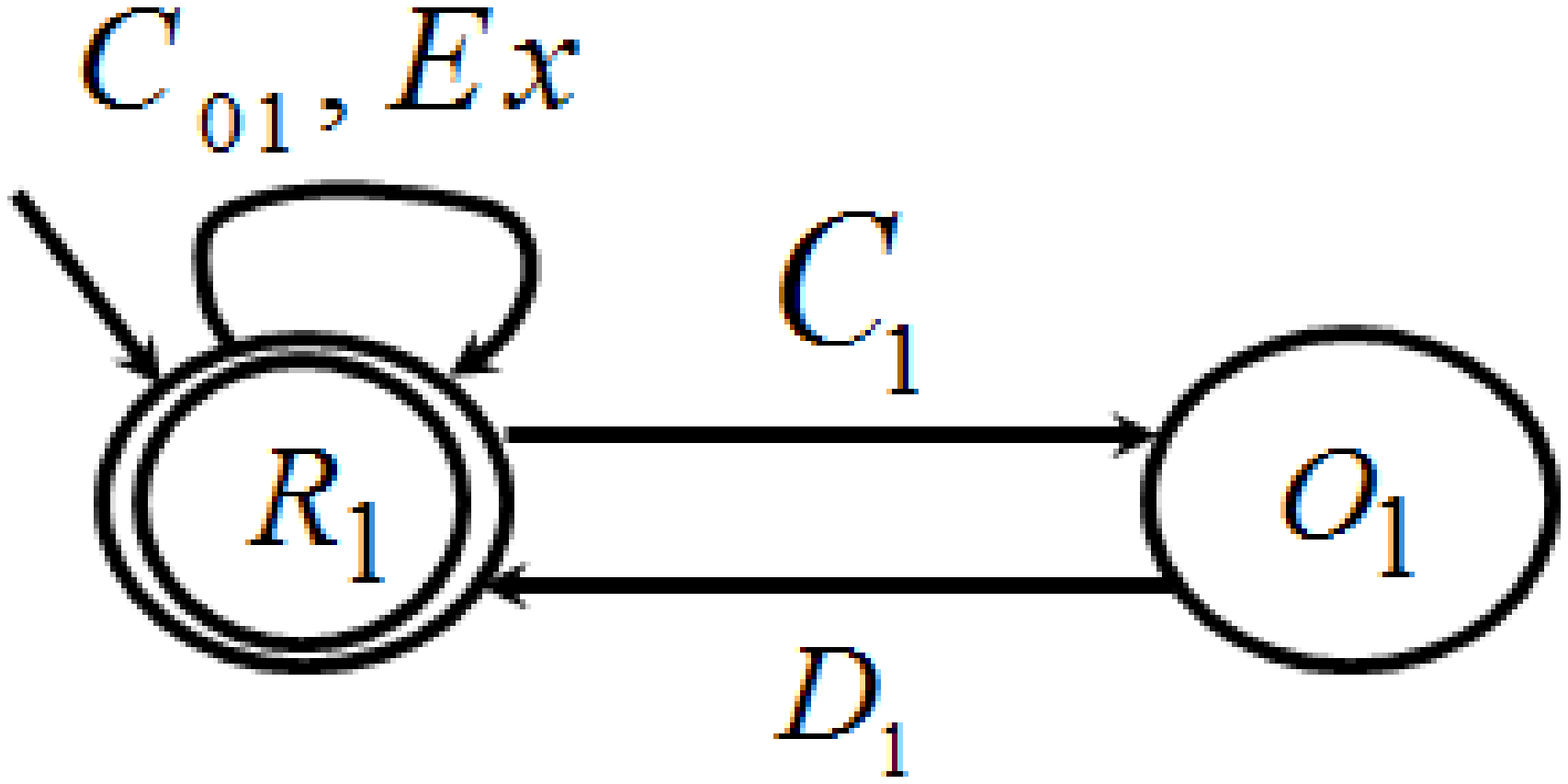}
  \label{UAV1DESmodel}
}
\subfigure[] 
{
  \includegraphics[width=1in, height=.75in]{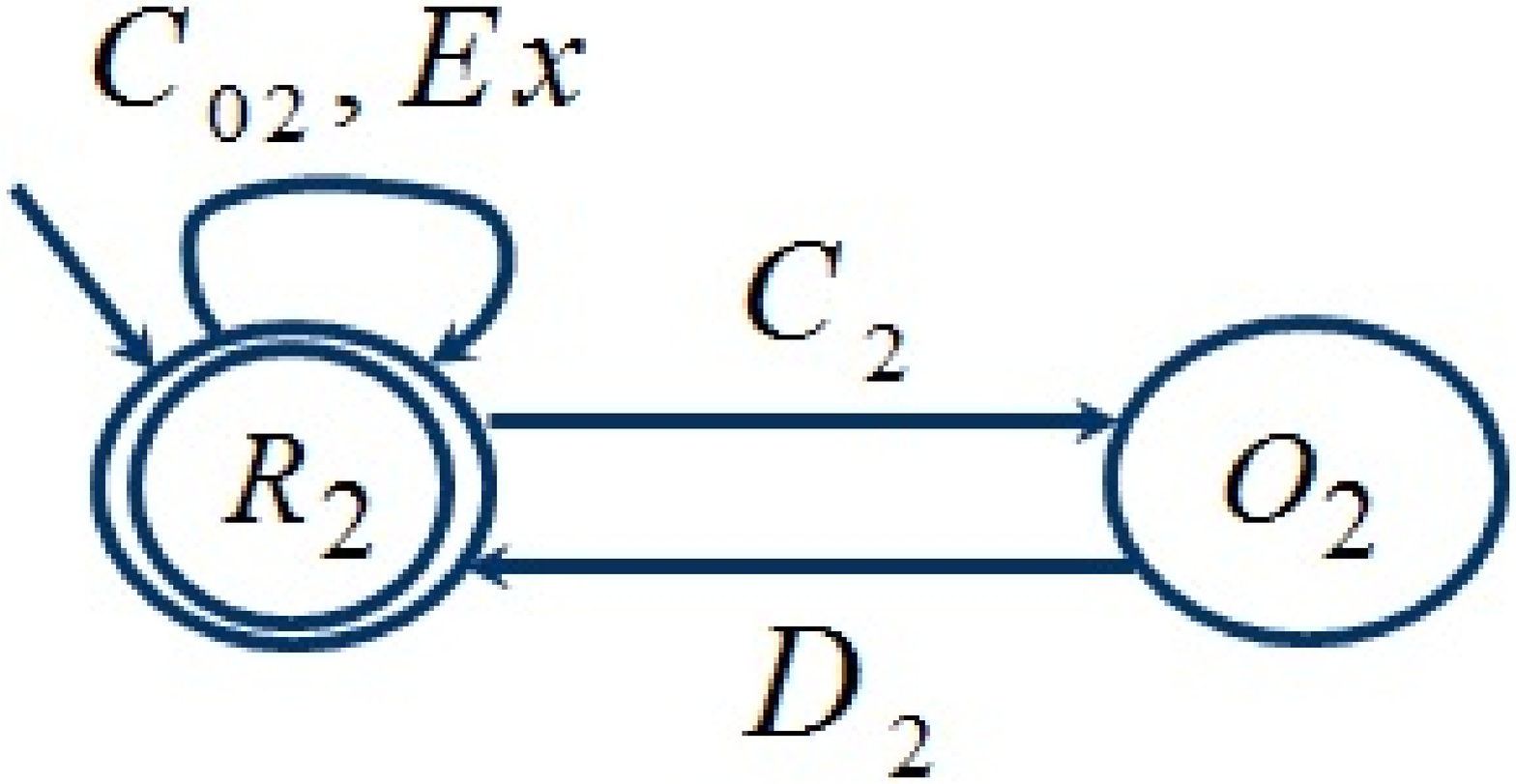}
  \label{UAV1DESmodel}
}
 \caption{(a) DES model of $UAV_1$. (b) DES model of $UAV_2$.} 
\label{UAVDESmodel}
\end{figure}

For $UAV_1$, the discrete model of the system over the partitioned space can be described by the automaton  $A_1=\left(Q_1, {q_0}_1, E_1, \delta_1,{Q_m}_1
\right)$ whose set of discrete states is $Q_1=\{R_1,O_1\}$, and its event set is ${E _1} =C_1\cup \{{C_0}_1\}\cup D_1 \cup Ex$, where $C_1=\{{C_r}_1^ + ,{C_r}_1^ - ,{C_{\theta }}_1^ + ,{C_{\theta }}_1^ -\}$ and $D_1= \{ {d_{i,j}}_1|\,\,1 \le i \le {n_r} - 1,1 \le j \le {n_\theta } - 1\}$. When $UAV_1$ is in one of the regions $R_{i,j}$, in the abstract model it is considered to be in the discrete state $R_1$. Then, one of the actuation commands belong to $C_1$ drives the UAV to one of its adjacent regions. In this case, right after issuing the actuation commands, the system transits to the detection state $O_1$ and waits until the UAV enters a new region. Crossing  boundaries of the new region, a detection event belonging to $D_1$ will be generated which shows the UAV has entered the new region $R_{i',j'}$. The command ${C_0}_1$, keeps the UAV in the current region and does not change the discrete state of the system. We use the notation ${D_{M}}_1=\{ {d_{i,j}}_1|\,\,1 \le i =1,1 \le j \le {n_\theta } - 1\}$ to denote the detection events, which show entering a region in the first circle, and $d_1=D_1- {D_{M}}_1=\{ {d_{i,j}}_1|\, 1 < i \le {n_r} - 1,1 \le j \le {n_\theta } - 1\}\subseteq D_1$ for the rest of detection events.  Here, $Ex=\{Ca_{12F},\,Ca_{12N},Ca_{21F},Ca_{21N},\,\,Stop_1,\,Stop_2,{R_{21}},\,{R_{12}}\}$ is the set of  external events which are required for the collision avoidance and do not change the state of the system. The events belong to $CA=\{Ca_{12F},\,Ca_{12N},Ca_{21F},Ca_{21N}\}$ show the collision alarms, in which the events  in $CA_1=\{C{a_{12F}}, C{a_{12N}}\}$ show that $UAV_2$ enters the alarm zone of $UAV_1$ and accordingly, the events in  $CA_2=\{C{a_{21F}},C{a_{21N}}\}$ show that  $UAV_1$ enters the alarm zone of $UAV_2$. The details will be discussed in Section \ref{Designing the supervisor for collision avoidance}. The events $Stop_1$ and $Stop_2$ are the commands that request $UAV_1$ and $UAV_2$ to stop at their current position in the relative frame and the command $R_{12}$ and $R_{21}$ release them, respectively.  Similar definitions can be given for the DES model of $UAV_2$. The graph representation of the discrete models of $UAV_1$ and $UAV_2$ are shown in Fig. \ref{UAVDESmodel}. In these graphs, the arrows starting from one
state and ending to another state represent the transitions, labeled
by the events belong to $E_i$.  The entering arrows stand for the
initial states. Marked states are shown by double circles.

In the DES model of $UAV_k$, $k=1,2$, the event set $E_k$ consists of  the controllable
event set ${E_c}_k=\{ {C_0}_k$, ${C_r}_k^ +$, ${C_r}_k^ -$, ${C_{\theta }}_k^ +$, ${C_{\theta }}_k^ -$, $Stop_1$, $Stop_2$, ${R_{21}}$, ${R_{12}}\}$ and the uncontrollable event set ${E_{uc}}_{k}=\{Ca_{12F},\,Ca_{12N},Ca_{21F},Ca_{21N}\}\cup D_1$. The uncontrollable events are those that cannot be
affected by the supervisor. A
language $K$ is controllable with respect to the language
$L(A)$ and the event set $E_{uc}$  if and only if
$\forall s \in K$ and $\sigma  \in E_{uc}$, if $s\sigma \in
L(A)$, then $s\sigma  \in K $.
Indeed, the controllability is the existence condition of a supervisor for the control goal described by the specification $K$ \cite{Cassandras2008}.

\section{Designing a decentralized modular supervisor  for the formation control of the UAVs} \label{Decentralized Modular supervisor design for the formation control of the UAVs}

Given the discrete model of follower UAVs over the partitioned space, it is possible to design the supervisor to achieve a desired order of events to accomplish the formation. Indeed, the
supervisor, $S$, observes the executed strings of the plant $A$ and
disables the undesirable controllable events. Here, we assume that
all of the events are observable. The generated language  and marked language
of the closed-loop system, $L(S/A)$ and $L_m(S/A)$, can be
constructed as follows:
\\(1) $\varepsilon  \in L(S/
  A)$ \\ (2) $\left[(s \in L(S/
  A))\,\,and\,\,(s\sigma
   \in L(A))\,\,and\,\,(\sigma  \in L( S))\right] \Leftrightarrow
 (s\sigma  \in  L(S/A)) $ \\ (3) $L_m(S/A)=L(S/A)\bigcap L_m(A)$\\
where $s$ is the string that  has been generated so far by the plant $A$, and $\sigma$
is an event, which the supervisor $S$ should decide whether  keep it
active or not in the supervised system $S/A$.

Within this framework one can use parallel composition to facilitate
the control synthesis. Parallel composition is a binary operation
between two automata which can be defined as follows:

\begin{definition} (Parallel Composition \cite{Cassandras2008}) \label{parallel
composition} Let $A_i=\left( Q_i,q_i^0,E_i,\delta _i,{Q_m}_i\right)$,
$i=1,2$, be automata. The parallel composition (synchronous
composition) of $A_1$ and $A_2$ is the automaton $A_1||A_2=(Q =
Q_1 \times Q_2, q_0 = (q_1^0, q_2^0), E = E_1 \cup E_2,
\delta, Q_m={Q_m}_1$  $\times {Q_m}_2)$, with $\delta$ defined as
 $\forall (q_1, q_2)\in Q, e\in E: \delta(\left(q_1, q_2),
   e\right)=\\
    \left\{
\begin{array}{ll}
    \left(\delta_1(q_1, e), \delta_2(q_2, e)\right), & \hbox{if $\delta _1(q_1,e)!, \delta _2(q_2,e)!,$}\\
    & \hbox{$e\in E_1 \cap E_2$;}\\
    \left(\delta_1(q_1, e), q_2\right), & \hbox{if $\delta _1(q_1,e)!, e\in E_1 \backslash E_2$;} \\
    \left(q_1, \delta_2(q_2, e)\right), & \hbox{if $\delta _2(q_2,e)!, e\in E_2 \backslash E_1$;} \\
     \hbox{undefined}, & \hbox{otherwise.}
\end{array}\right.$
\end{definition}

Here, the parallel composition is used to
combine the plant's discrete
model and the supervisor as follows:

\begin{lemma}\label{supervisor} \cite{KumGar:95} Let $A=(Q,\,q_0,\,E,\,\alpha,\,Q_m)$, be the plant automaton
 and $K\subseteq E ^*$ be the
desired marked language. There exists a nonblocking supervisor $S$ such
that $L_m(S/A)=L_m(S\|A)=K$ if $\emptyset \neq K=
\bar{K}\bigcap L_m(A)$ and $K$ is controllable.  In this case, $S$
could be any automaton with $L(S)=L_m(S)=\bar{K}$.
\end{lemma}

Now, using the above lemma, it is possible to design the supervisor for the formation problem described in Problem \ref{problem1}, which includes two modules: 1- Reaching and keeping the formation and 2- Avoiding collision. Next lemma describes how to design the supervisors in a modular way.

\begin{lemma}\label{modularsupervisor} \cite{KumGar:95} Let $A=(Q,\,q_0,\,E,\,\alpha,\,Q_m)$ be the plant automaton
 and the prefix-closed controllable languages $K_1, K_2\subseteq E ^*$ be the
desired marked specifications. Suppose there exist nonblocking supervisors ${S}_1$ and ${S}_2$  such that  $L_m({S}_1/A)=L_m({S}_1\|A)=K_1$ and $L_m({S}_2/A)=L_m({S}_2\|A)=K_2$, then  $S={S}_1\|{S}_2$ is a nonblocking supervisor with $L_m(S\|A)=K_1\bigcap K_2$. $\blacksquare$
\end{lemma}

\subsection{Designing the supervisor for reaching and keeping the formation}
\begin{figure}
\centering 
\subfigure[] 
{
  \includegraphics[width=2 in, height=1.5in]{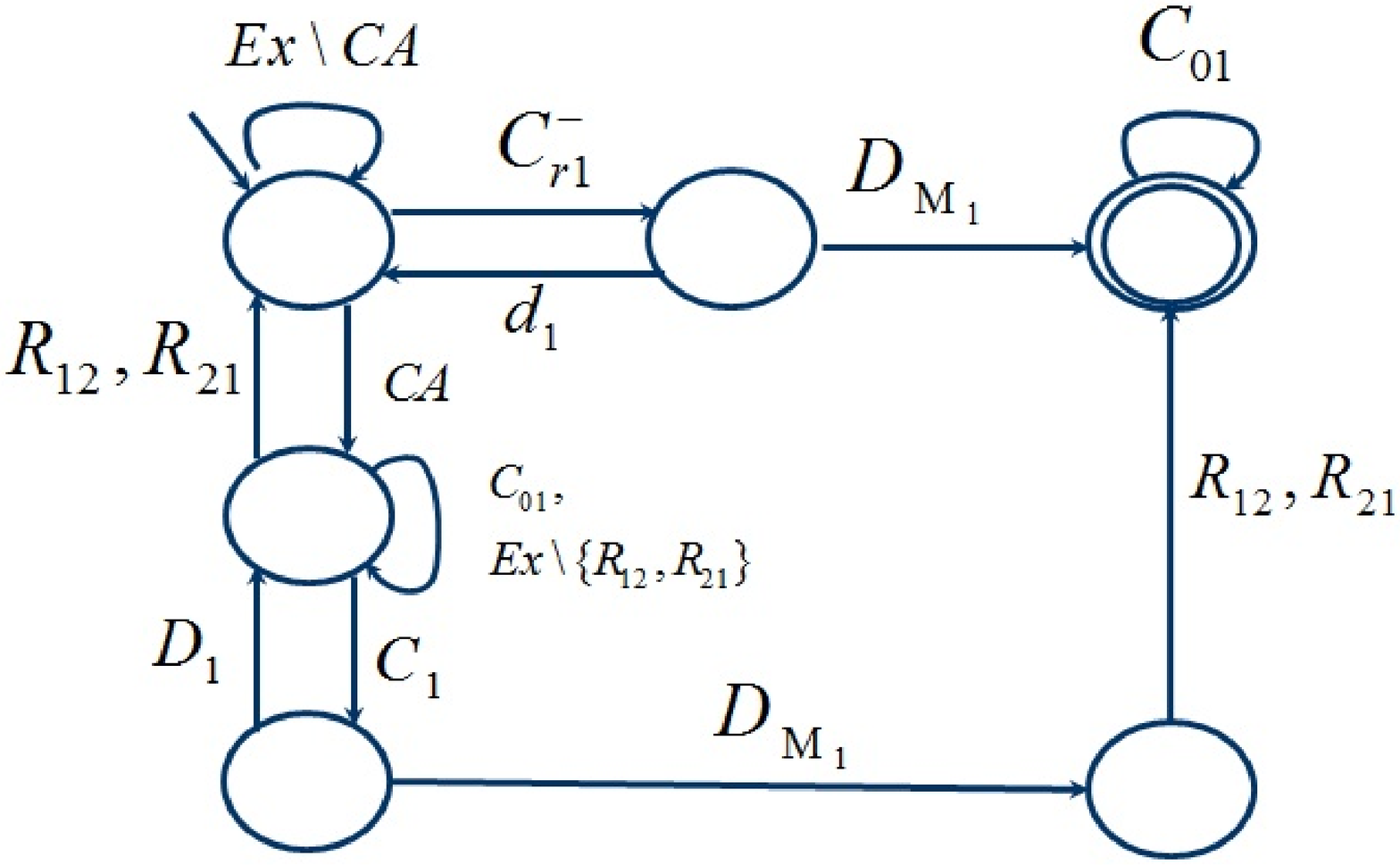}
  \label{ReachingFormation1}
}
\subfigure[] 
{
  \includegraphics[width=2in, height=1.5in]{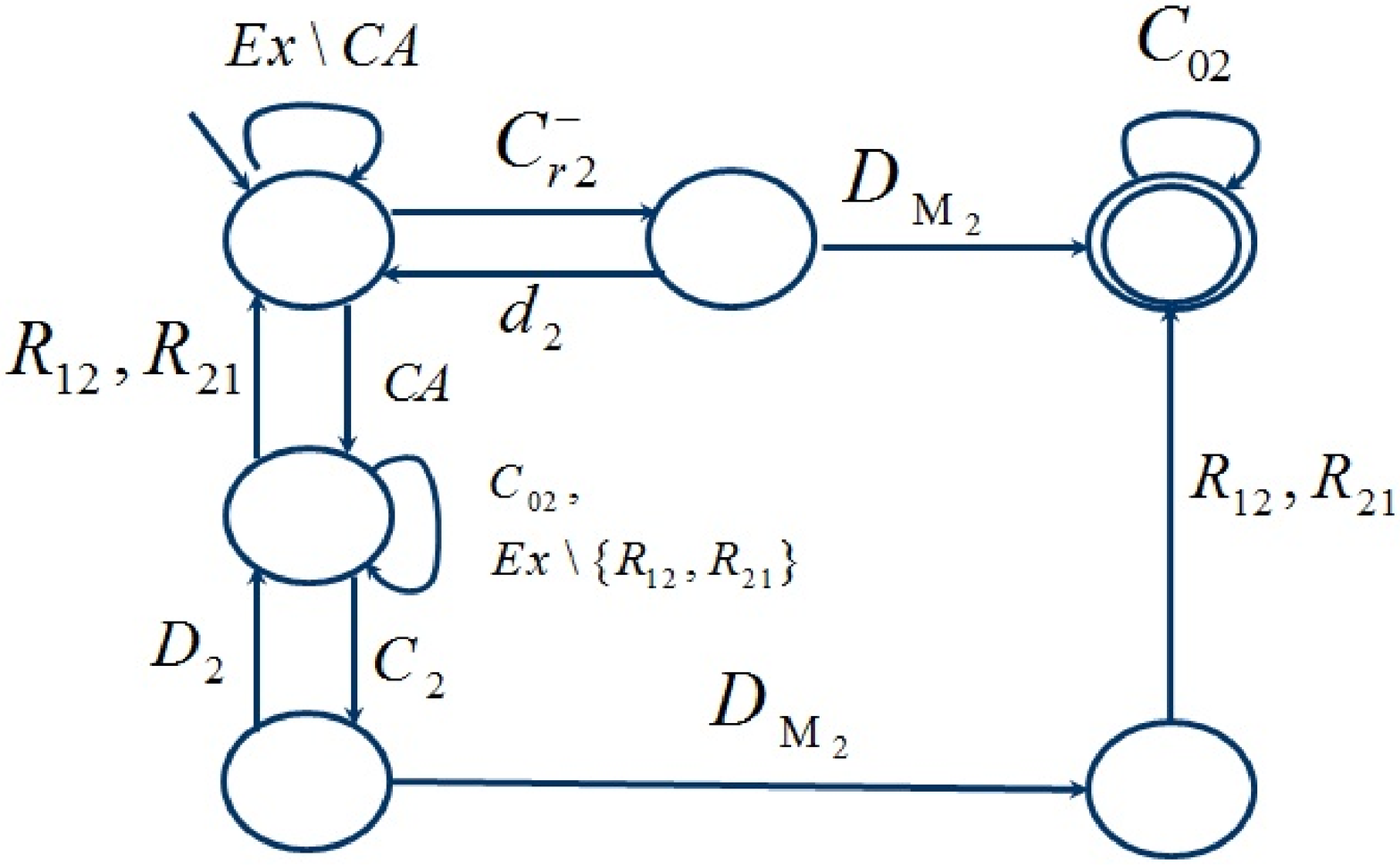}
  \label{ReachingFormation2}
}
 \caption{(a) The specification for reaching and keeping the formation for UAV1. (b) The specification for reaching and keeping the formation for UAV2.} 
\label{ReachingFormation}
\end{figure}

For reaching the formation, it is sufficient to directly drive each of the follower
UAVs  towards  one of  the regions $R_{1,j}$, $1\leq j \leq
n_\theta -1$,  located in the first
circle in their corresponding partitioned motion space. After reaching $R_{1,j}$, the UAVs should remain inside it,
to keep the formation.  The specifications ${K_F}_1$ and ${K_F}_2$ for reaching and keeping the specification for $UAV_1$ and $UAV_2$ are realized in Fig.
\ref{ReachingFormation}. 
When the k'th follower UAV is not in the first circle, the command ${C_r}_k^-$ will be
generated to push the UAV towards the origin. Entering a new
region, one of the events from   $d_k=\{ {d_{i,j}}_k|\,\,1 < i \le {n_r} - 1,1 \le j \le {n_\theta } - 1\}$ will appear.  This will continue until one of the events from ${D_{M}}_k=\{ {d_{i,j}}_k|\,\,i=1,1 \le j \le {n_\theta } - 1\}$ be generated, which shows that the
formation is reached.   In this case, the event ${C_0}_k$ is
activated, which keeps the system trajectory inside the first region. If  a collision alarm happens to $UAV_k$,  the formation supervisor does not change the generable language after the events belonging to  $CA$, and lets the collision avoidance supervisor  handle it  until  the collision be avoided and the UAV be released to resume the formation task.

It can be seen that ${K_F}_k$, $k=1,2$ are controllable with respect to the plant language ${L(A_k)}$ and the event set ${E_{uc}}_k$,  as they do not
disable any uncontrollable event. Therefore, based on Lemma  \ref{supervisor}, there exist  supervisors that can
control the plants $A_1$ and $A_2$ to achieve these specifications. The supervisors are
the realization of the above specifications  in which all states are
marked. Marking all states of the supervisors allows the closed-loop
marked states  to be solely determined by the plants' marked states.
The supervisor for  reaching the formation and keeping the formation of $UAV_k$
is denoted by ${A_F}_k$. 


\vspace{10 mm}

\subsection{Designing the supervisor for collision avoidance}\label{Designing the supervisor for collision avoidance}
\begin{figure}[ihtp]
      \begin{center}
     \includegraphics[width=.2\textwidth]{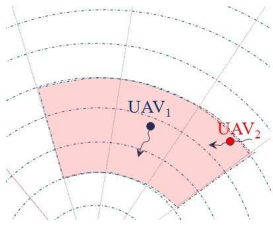}
        \caption{$UAV_2$ enters the alarm zone of $UAV_2$.}
        \label{collision}
        \end{center}
 \end{figure}
\begin{figure*}[ihtp]
      \begin{center}
     \includegraphics[width=.6\textwidth]{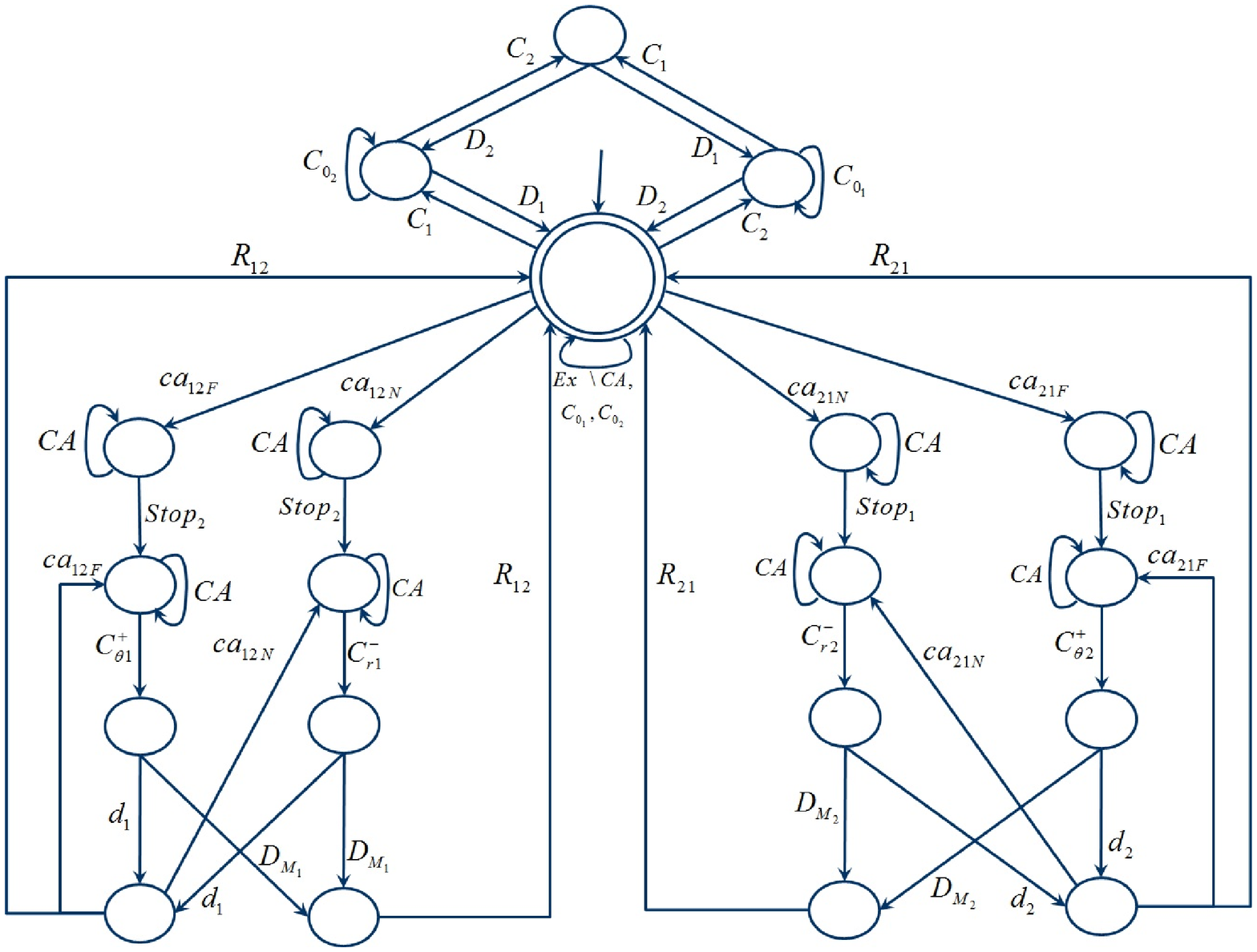}
        \caption{The specification for cooperative collision avoidance.}
        \label{cooperativeCollisionavoidance}
        \end{center}
 \end{figure*}

When $UAV_1$ is going to reach  its desired
position, in some situations, the other follower, $UAV_2$, may enter the alarm zone of $UAV_1$ (Fig. \ref{collision}), which requires these
UAVs to cooperatively avoid the collision. For this purpose, first, $UAV_1$ asks $UAV_2$ to stop in the relative frame and then, $UAV_1$ finds a path to safely get away from $UAV_2$. After avoiding the collision, $UAV_1$ releases $UAV_2$ and both UAVs resume their normal operation for reaching the formation. Similar strategy is taken when $UAV_1$ enters the alarm zone of $UAV_2$. This specification, $K_C$, is shown in Fig. \ref{cooperativeCollisionavoidance} whose left side shows that after appearing one of the events ${ca_{12}}_F$ or  ${ca_{12}}_N$, $UAV_1$ realizes that $UAV_2$ has entered its alarm zone. Therefore, by event $Stop_2$, $UAV_1$ requests $UAV_2$ to stop for a while to safely manage the situation.  The event ${ca_{12}}_F$ shows that $UAV_2$ is in front of the path of $UAV_1$ towards its destination and hence, to avoid the collision it is sufficient that $UAV_1$  turns anticlockwise to change its azimuth angle, $\theta$,  by
activating the command $C_{\theta} ^+$. This will continue until removing the collision alarm. Then, $UAV_1$ releases $UAV_2$, and reaching the formation can be resumed by the reaching formation supervisor which was explained in the previous section. Meanwhile, if $UAV_1$ enters one of the regions in the first circle, one of the events belong to ${D_{M}}_1$ appears which means that $UAV_1$ has reached its desired formation and should remain there for the rest of mission. Similarly, the right side of Fig. \ref{cooperativeCollisionavoidance}, shows the collision avoidance mechanism when $UAV_1$ enters the alarm zone of $UAV_2$. If neither of collision avoidance alarms from the set $CA$ happens, then $UAV_1$ and $UAV_2$ can do their normal operations by independent enabling of events $C_1$ and $C_2$ followed by the detection signals $D_1$ and $D_2$ in any order as shown on the top of Fig. \ref{cooperativeCollisionavoidance}. The other module, the reaching formation supervisor, will manage this situation.

It can be verified  that ${K_C}$ is controllable with respect to the  language ${L(A_1|| A_2)}$ and the event set ${E_{uc}}_1 \cup {E_{uc}}_2$. Therefore, based on Lemma  \ref{supervisor}, there exists a  supervisor $A_c$ that can
control the plants $A_1$ and $A_2$ to achieve this joint specification. The supervisor is
the realization of the specification $K_C$  in which all states are
marked.

The collision avoidance supervisor, $A_C$, is a centralized supervisor which manages both  $UAV_1$ and $UAV_2$. To make this supervisor decentralized and to achieve local supervisors, we will utilize our proposed decomposition scheme introduced in \cite{karimadini2011guaranteed}.  Here, local supervisors can be achieved by the projection of the global supervisor to each agent's local event set. The projection of the global supervisor $A_C$ to the event set of $UAV_i$, $E_i$, is denoted by
$P_{E_i}(A_C)$, and  can be obtained  by replacing the events that
belong to $E\backslash E_i$ by $\varepsilon$-moves, and then,
merging the $\varepsilon$-related states.

Once the local supervisor automata are derived through the natural projection, the decentralized supervisor is then obtained using the parallel composition of local supervisor automata. Parallel composition captures the logical behavior of concurrent distributed systems by allowing each subsystem to evolve individually on its private events, while synchronize with its neighbors on shared events for cooperative tasks.

The obtained decentralized supervisor is then compared with the original global supervisor automaton using the bisimulation relation.
\begin{definition}
\label{simulation}
Consider two automata $A_i=( Q_i, q_i^0$, $E,
\delta _i)$, $i=1, 2$.
The automaton $A_1$ is said to be similar to $A_2$ (or $A_2$ simulates $A_1$), denoted
by $A_1\prec A_2$, if there exists a relation $R$ from $A_1$ to $A_2$ over $Q_1$, $Q_2$ and with respect to $E$, such that (1) $(q_1^0, q_2^0) \in R$, and (2) $\forall\left( {q_1 ,q_2 } \right) \in R,
q'_1 \in \delta_1(q_1, e)$, then $\exists q_2^{\prime}\in Q_2$ such
that $q'_2 \in \delta_2(q_2, e)$, $\left( {q'_1 ,q'_2 } \right) \in
R$. Automata $A_1$ and $A_2$ are said to be bisimilar (bisimulate each
other), denoted by $A_1\cong A_2$ if $A_1\prec A_2$ with a simulation relation $R_1$, $A_2\prec A_1$
with a simulation relation $R_2$ and $R_1^{-1} = R_2$, where $R_1^{-1}= \{(y,x)\in Q_2\times Q_1|(x,y)\in R_1\}$.
\end{definition}


Based on these definitions we can formally describe the decomposability conditions with respect to two local event sets.
\begin{lemma} (Theorem $4$ in \cite{karimadini2011guaranteed}) \label{Task Automaton
Decomposition} A deterministic  automaton $A = (Q, q_0, E=E_1\cup
E_2, \delta)$ is decomposable with respect to parallel composition
and natural projections $P_i$, $i=1,2$, such that
$A\cong P_1(A)||P_2(A)$ if and only if $A$ satisfies the
following decomposability conditions (DC): $\forall e_1 \in
E_1\backslash E_2, e_2 \in E_2\backslash E_1, q\in Q$, $s\in E^*$,
\begin{itemize}\item $DC1$: $[\delta(q,e_1)!\wedge
\delta(q,e_2)!]\Rightarrow [\delta(q, e_1e_2)! \wedge \delta(q,
e_2e_1)!]$;
\item $DC2$: $\delta(q, e_1e_2s)!\Leftrightarrow \delta(q, e_2e_1s)!$;
 \item $DC3$:
$\forall s, s^{\prime} \in E^*$, $s\neq s^{\prime}$, $p_{E_1\cap
E_2}(s)$, $p_{E_1\cap E_2}(s^{\prime})$ start with the same common
event $a\in E_1 \cap E_2$, $q\in Q$: $\delta(q, s)! \wedge \delta(q,
s^{\prime})! \Rightarrow \delta(q,
\overline{p_1(s)}|\overline{p_2(s^{\prime})})! \wedge \delta(q,
\overline{p_1(s^{\prime})}|\overline{p_2(s)})!$;
\item $DC4$: $\forall i\in\{1, 2\}$, $x, x_1, x_2 \in Q_i$, $x_1\neq x_2$,
$e\in E_i$, $t\in E_i^*$, $x_1\in\delta_i (x, e)$, $x_2\in\delta_i
(x, e)$: $\delta_i (x_1, t)! \Leftrightarrow \delta_i(x_2, t)!$.
\end{itemize}
\end{lemma}

where, $\bar{K}=\{s\in \Sigma ^*|(\exists t \in \Sigma ^* ) st \in K \}$ is the prefix closure of the language $K$. The decomposability conditions $DC1$ and $DC2$ respectively guarantee that any decision on the selection or order of two transitions can be done by the team of agent, while conditions $DC3$ and $DC4$ respectively ensure that the interaction of local automata $P_1(A)$ and $P_2(A)$ neither allows an illegal string that is not in $A$, nor stops a legal string of $A$.

Now assume that given the global task and local plants, a global supervisor is designed and decomposed into local supervisors such that each closed loop system (the supervised local plant with the corresponding local controller) satisfies the global task. In this decentralized cooperative control architecture we are then interested to check whether the entire system satisfied the global task.

\begin{problem}(Decentralized cooperative control problem)\label{Decentralized cooperative control problem}
Consider a plant, represented by a parallel distributed system
$A_P:=\overset{2}{\underset{i=1}{\parallel} }A_{P_i}$, with local
event sets $E_i$, $i=1,2$, and let the global specification is
given by a deterministic task automaton $A_S$ over
$E=\overset{2}{\underset{i=1}{\cup} }E_i$.
Furthermore, suppose that there exist a decomposable deterministic global
controller automaton $A_C \cong \overset{2}{\underset{i=1}{\parallel} }P_i(A_C)$, so that $A_P\parallel A_C \cong A_S$.
Then, whether the local controllers can lead the team to satisfy the global
specification in a decentralized architecture,
$\overset{2}{\underset{i=1}{\parallel}}(A_{P_i}\parallel
P_i(A_C))\cong A_S$.
\end{problem}

Following result considers a team of two local plants and introduces the supervisor decomposability and satisfaction of the global task by each local supervised plant as a sufficient condition for the satisfaction of global task by the team.


\begin{theorem}(Decentralized cooperative control using supervisor decomposition)\label{Decentralized cooperative control using supervisor decomposition}
Consider a plant, represented by a parallel distributed system
$A_{P_1}\parallel A_{P_2}$, with local
event sets $E_i$, $i=1, 2$, and let the global specification is
given by a  task automaton $A_S$ over
$E= E_1 \cup E_2$.
Furthermore, suppose that there exist a deterministic global
controller automaton $A_C \cong P_1(A_C) \parallel P_2(A_C)$, so that $A_C \parallel A_p \cong A_S$.
Then, the entire closed
loop system satisfies the global
specification, in the sense of
bisimilarity, i.e.,
$\overset{2}{\underset{i=1}{\parallel}}(A_{P_i}\parallel
P_i(A_C))\cong A_S$, provided the decomposability conditions $DC1$, $DC2$, $DC3$ and $DC4$ for $A_C$.
\end{theorem}

%

\begin{figure}
\centering 
\subfigure[] 
{
  \includegraphics[width=3.1 in, height=3in]{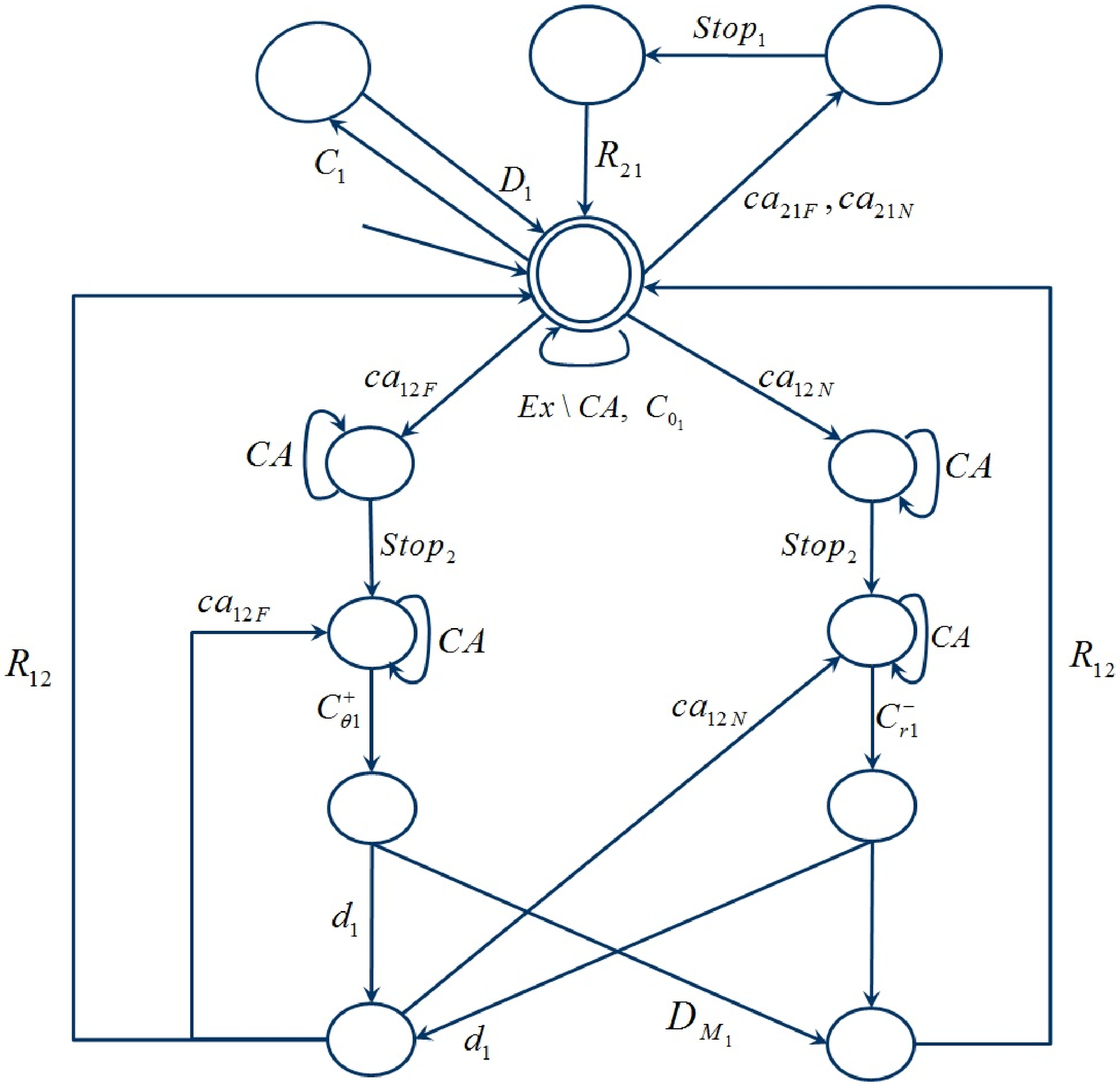}
  \label{SupervisorCollisionavoidance1}
}
\subfigure[] 
{
  \includegraphics[width=3.1in, height=3in]{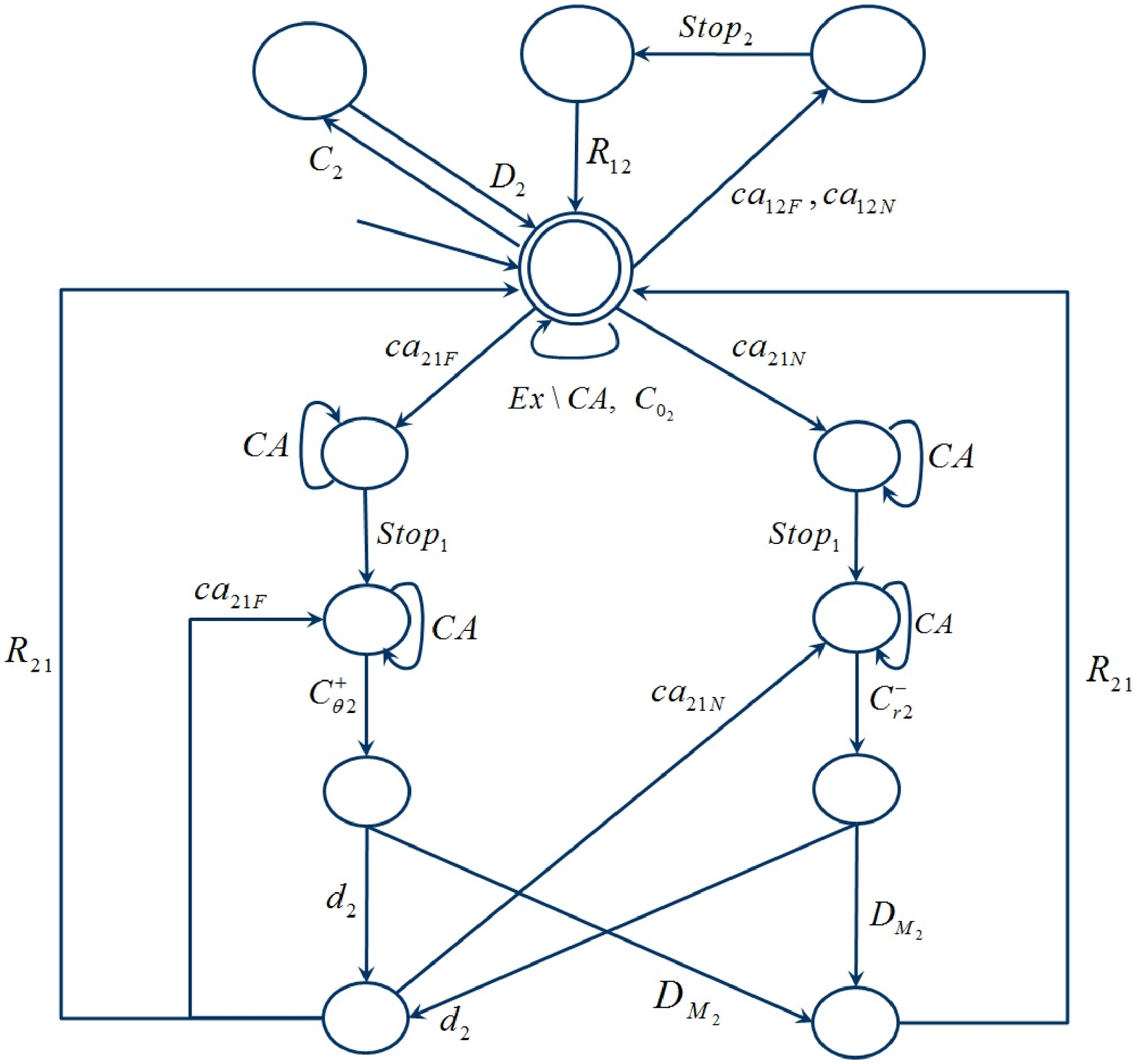}
  \label{SupervisorCollisionavoidance2}
}
 \caption{(a) The local supervisor for collision avoidance for $UAV_1$. (b) The local supervisor for collision avoidance for $UAV_2$.} 
\label{decomposedSupervisorCollisionavoidance}
\end{figure}

The significance of this result is the decentralized implementation of the global supervisor, $A_C$,  given in Fig. \ref{cooperativeCollisionavoidance}, by decomposing $A_C$, into local supervisors. As it can be seen in $A_C$, the successive  and adjacent events from pairs of private event sets (from different local event sets) $({C_0}_1, C_2)$, $({C_0}_1,D_2)$, $({C_0}_2,C_1)$, $({C_0}_2,D_1)$, $(C_1,D_2)$, $(C_2,D_2)$, appear in both orders in the global supervisor automaton therefore $DC1$ and $DC2$ are satisfied. Moreover, among common events $R_{12}$, $R_{21}$, $CA_1=\{ca_{12F},ca_{12N}\}$,  $CA_2=\{ca_{21F},ca_{21N}\}$, $Stop_1$, and $Stop_2$, the events $R_{12}$, $R_{21}$, $Stop_1$, and $Stop_2$ are not shared between different strings. Strings just share the  events $CA_1$,  $CA_2$, where the corresponding local strings do not interleave on these events because of predecessor common events before $CA_1$,  $CA_2$. Therefore $DC3$ also is fulfilled. Finally, $DC4$ is satisfied because of the determinism of local automata $P_1(A_C)$ and $P_2(A_C)$, and hence, the supervisor automaton $A_C$ is decomposable into ${A_C}_1=P_1(A_C)$ and ${A_C}_2=P_2(A_C)$, shown in Fig. \ref{decomposedSupervisorCollisionavoidance}, so that ${A_C}_1\parallel {A_C}_2 \cong A_C$.



\section{Verifying the algorithm through a hardware-in-the-loop simulation platform}\label{Hardware-in_the_loop Simulation results}
\begin{figure}[ihtp]
      \begin{center}
     \includegraphics[width=.25\textwidth]{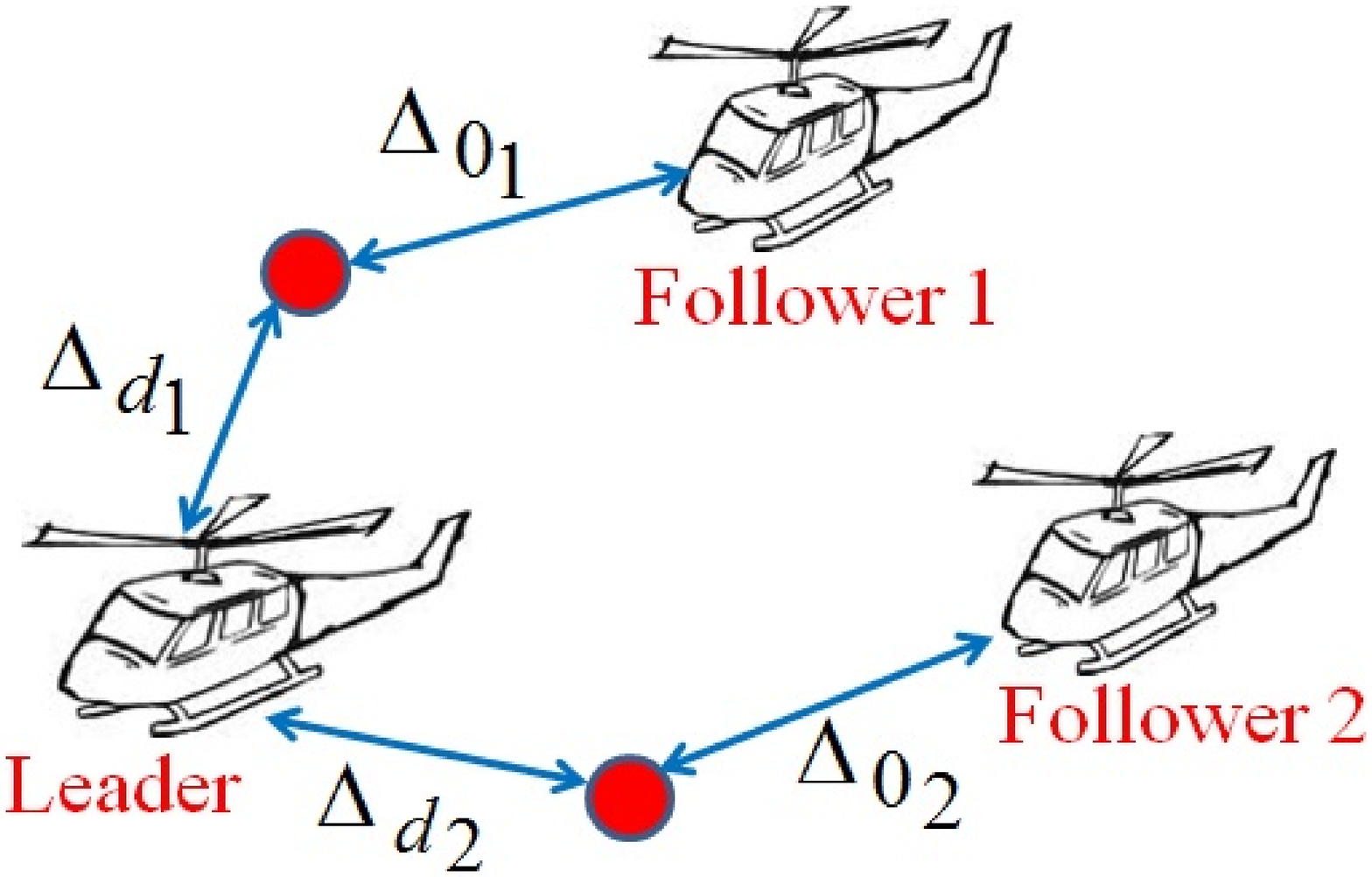}
        \caption{The schematic of a formation scenario with two followers and one leader}
        \label{schematic}
        \end{center}
 \end{figure}

To verify the proposed algorithm,  we have used a
hardware-in-the-loop simulation platform \cite{Cai20091057}
developed for NUS UAV helicopters \cite{Peng20092333}. In this
platform, the nonlinear dynamics of the UAVs  have been replaced
with their nonlinear model, and all software and hardware components
that are involved in a real flight test remain active during the
simulation so that the simulation results achieved from this
simulator are very close to the actual flight tests. This multi-UAV
simulator test bed is used to verify the proposed algorithm. For this purpose, consider two followers that should track a leader UAV with a desired distance, as shown in Fig. \ref{schematic}. The distance between the desired position of the $Follower_1$ and $Follower_2$ and the leader UAV are
${\triangle_d}_1=(12,10)$ and ${\triangle_d}_2=(-12,-10)$, respectively. The follower UAVs initially are not at the desired position. The initial distance between $Follower_1$ and its desired position is  ${\triangle_0}_1=(-41.9, -0.9)$, and the initial distance between $Follower_2$ and its desired position is ${\triangle_0}_2=(-17.5,0.5)$. $Follower_1$ after 34.8 sec and $Follower_2$ after 14.3 sec reach the formation and then, they will keep the formation.

 \begin{figure*}[ihtp]
      \begin{center}
     \includegraphics[width=1\textwidth]{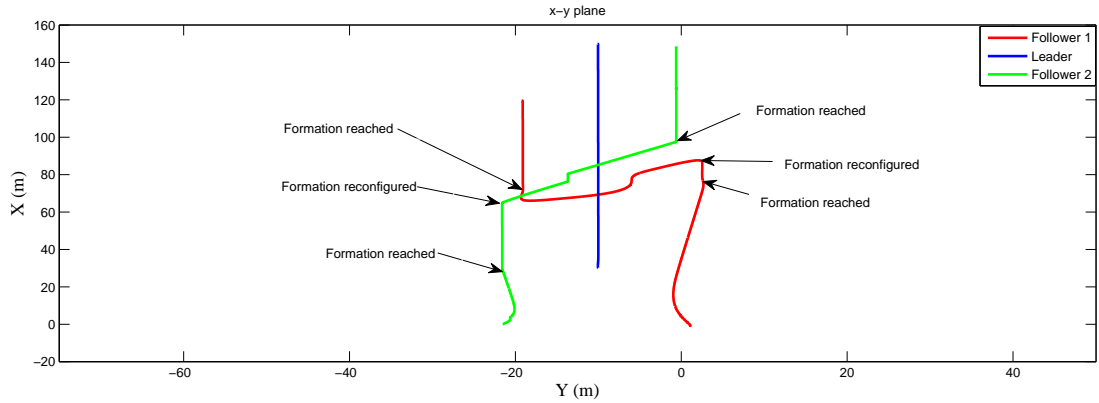}
        \caption{The position of the UAVs in the x-y plane.}
        \label{xy}
        \end{center}
 \end{figure*}

 \begin{figure*}[ihtp]
      \begin{center}
     \includegraphics[width=1\textwidth]{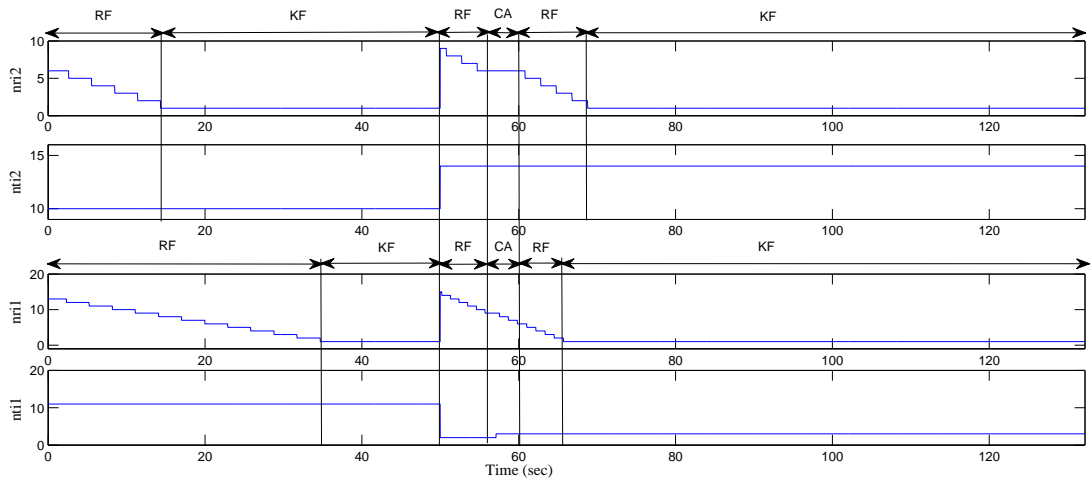}
        \caption{The indices of $\theta$ and $r$ for the traversed regions by $Follower_1$ and $Follower_2$.}
        \label{ntriboth2}
        \end{center}
 \end{figure*}

 After 50 sec, the formation switches. For the new formation, the desired distance of the followers  from the leader are ${\triangle_d}_1=(-30,-10)$ and ${\triangle_d}_2=(0,10)$, while their initial distances from the desired position are ${\triangle_0}_1=(40.5,23.3)$ and ${\triangle_0}_2=(-14.5,-23)$. When the followers are trying to reach the desired formation, at $t=55.8$ $sec$, $Follower_2$ enters the alarm zone of $Follower_1$. As described in Section \ref{Designing the supervisor for collision avoidance}, to avoid collision, $Follower_1$ asks $Follower_2$ to stop in the relative frame, and then it turns to handle the situation. After removing the collision alarm, both followers have resumed their normal operation to reach and keep the formation. The indices of the traversed regions for $\theta$ and $r$ are shown in Fig. \ref{ntriboth2}.
   The position of the UAVs in x-y plane is shown in Fig. \ref{xy}.

\section{CONCLUSION}\label{CONCLUSION}
In this paper,  a collision free formation control algorithm  was proposed using hybrid supervisory control techniques. The proposed supervisor has a modular structure and can accomplish three main tasks: reaching the formation, keeping the formation, and collision avoidance. This control structure was implemented decentralizedly so that local (decomposed) supervisors can treat the distributed agents to achieve a globally safe and collision free environment.The efficiency of the proposed approach was verified through hardware-in-loop simulation results.


%
%
%
%

\section*{ACKNOWLEDGMENT}
The financial supports from NSF-CNS-1239222 and NSF- EECS-1253488 for this work are greatly acknowledged.

\bibliographystyle{elsarticle-num}
\bibliography{bibhybrid3d}


\end{document}